\documentclass[journal]{IEEEtran}
\usepackage{cite,amsmath,graphicx,bbm}
\usepackage{algorithm}
\usepackage{subcaption}
\usepackage{caption}
\usepackage{url}
\usepackage{booktabs}
\usepackage{algpseudocode}
\usepackage{hyperref}
\usepackage{multicol}
\usepackage{amssymb}
\usepackage[table]{xcolor}

\usepackage{letltxmacro}
\LetLtxMacro{\originaleqref}{\eqref}
\renewcommand{\eqref}{Eq.~\originaleqref}

\usepackage{color}
\title{Depth Map Estimation of Dynamic Scenes Using Prior Depth Information}

\author{James~Noraky,~\IEEEmembership{Student Member,~IEEE,}
	Vivienne~Sze,~\IEEEmembership{Senior Member,~IEEE}}

\begin{document}

\maketitle

\begin{abstract}
Depth information is useful for many applications. Active depth sensors are appealing because they obtain dense and accurate depth maps. However, due to issues that range from power constraints to multi-sensor interference, these sensors cannot always be continuously used. To overcome this limitation, we propose an algorithm that estimates depth maps using concurrently collected images and a previously measured depth map for dynamic scenes, where both the camera and objects in the scene may be independently moving.  To estimate depth in these scenarios, our algorithm models the dynamic scene motion using independent and rigid motions. It then uses the previous depth map to efficiently estimate these rigid motions and obtain a new depth map. Our goal is to balance the acquisition of depth between the active depth sensor and computation, without incurring a large computational cost. Thus, we leverage the prior depth information to avoid computationally expensive operations like dense optical flow estimation or segmentation used in similar approaches. Our approach can obtain dense depth maps at up to real-time (30 FPS) on a standard laptop computer, which is orders of magnitude faster than similar approaches. When evaluated using RGB-D datasets of various dynamic scenes, our approach estimates depth maps with a mean relative error of 2.5\% while reducing the active depth sensor usage by over 90\%.  
\end{abstract}


\section{Introduction}\label{sec:intro}
Depth information is useful for many applications that include robotics, augmented reality, and manufacturing. In order to obtain accurate depth measurements, active depth sensors like time-of-flight or structured light cameras are often used. These sensors obtain depth measurements in the form of a depth map, which is an image whose pixels represent the distance from the sensor to various points in the scene. However, for many applications, it is often undesirable to continuously use the active depth sensors to measure depth. Their high power consumption can reduce battery life as well as increase the heat dissipation, which distorts the depth measurements \cite{Sarbolandi2015}. Furthermore, in settings where many active depth sensors can interfere, schemes that mitigate interference often limit the rate at which depth maps can be measured altogether \cite{Buttgen2007,Buttgen2008,Li2015}. This motivates the need of estimating dense and accurate depth maps without using the active depth sensor all of the time. Additionally, the estimated depth maps must be causal and obtained with minimal latency and high throughput to support applications, like robotic navigation, that use depth to interact with its surrounding environment.    

One way to tackle this problem is to leverage concurrently collected images to estimate new depth maps without using the active depth sensor. Images are typically collected in many applications and are freely available for the purpose of depth estimation. The idea of using images to estimate depth has been explored in many applications, and many approaches exploit the temporal correlation across consecutive frames to estimate depth. However, many of these approaches are unappealing to use because they estimate depth with high latency and high computational costs. This is because estimating depth using consecutive images is inherently underdetermined, where changes in depth are 3D in nature but are  captured as 2D image displacements. To overcome this challenge, many of these techniques require both dense optical flow and segmentation, which are computationally expensive to obtain \cite{Fortun2015,Yao2019}. Consequently, we take a different approach to estimate depth to avoid incurring a large computational cost.
 
\begin{figure}[t]
   \centering
   \includegraphics{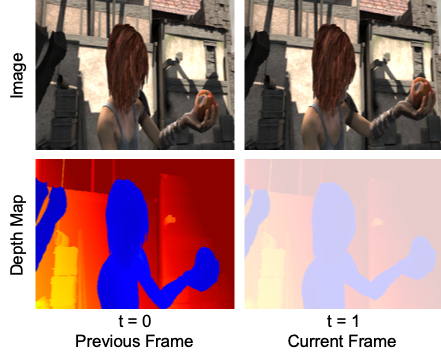}
   \caption{\textbf{Depth Estimation Setup}: The inputs to our algorithm are two consecutive and concurrently collected images and a previous depth map. The algorithm then estimates the current depth map. Here, $t$ denotes time. Because the previous depth map can either be measured or estimated, our technique can be used to sequentially estimate depth to further reduce the usage of the active depth sensor. In this scenario, we first estimate a new depth map using a previously measured depth map and then estimate subsequent depth maps using the previously \emph{estimated} depth map.}
   \label{fig:setup} 
\end{figure}

\begin{figure*}[ht]
   \centering
   \includegraphics{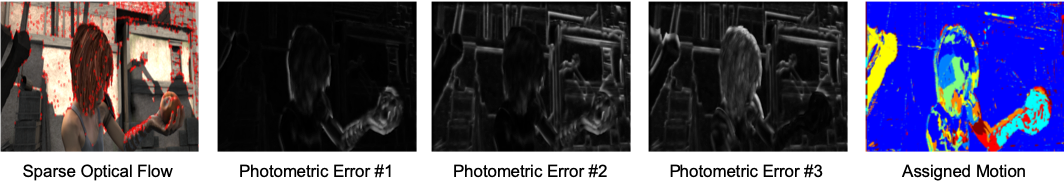}
   \caption{\textbf{Our Contributions}: We estimate depth maps efficiently by first estimating the independent rigid motions in the scene using the optical flow at a {sparse} set of pixels (shown in red). We then use each of the estimated rigid motions along with the previous depth map to reproject the previous image and obtain the photometric error. We show three examples of the photometric error, which can be efficiently computed. We then use these errors to assign the estimated rigid motions to the pixels of the previous depth map as shown in the last image, where the different colors represent the different assigned motions. As a consequence of our design decisions, our technique is able to estimate depth maps in dynamic scenes without dense optical flow or segmentation.}
   \label{fig:contributions} 
\end{figure*}

As shown in Figure \ref{fig:setup}, our algorithm estimates a new depth map using two consecutive and concurrently collected images and a \emph{previous} depth map. However, even with this additional prior information, estimating depth using consecutive images is still underdetermined. Here, we show how we overcome this challenge by using the prior information to efficiently estimate depth in dynamic scenes, where both the camera and objects in the scene can have independent and non-rigid motions, \emph{without} dense optical flow or prior segmentation. Previously, we used this setup to estimate depth maps for rigid objects and scenes \cite{Noraky2019}, which is ideal for applications, like SLAM, that assume a static environment. In this work, we increase the variety of scenes and applications that our framework can support. 

To estimate depth in dynamic scenes, we first model its motion using independent rigid motions. We then use the prior depth information to both estimate these rigid motions and assign them to the pixels of the previous depth map in order to estimate a new one. As shown in Figure \ref{fig:contributions}, the key contributions are: 

\begin{itemize}
   \item We show that the rigid motions in the scene can be estimated using the optical flow at a \emph{sparse} set of pixels without prior knowledge of the number of these motions. This allows us to avoid dense optical flow estimation, which is computationally expensive, and thus increase the throughput at which we estimate depth.
   \item We show that the estimated rigid motions can be efficiently assigned to the pixels of the previous depth map by using the photometric error between the current image and an image obtained by reprojecting the previous image using the estimated rigid motions and the previous depth map. Computing the photometric error is efficient, as its complexity is linear with respect to the number of pixels in the previous image, and it allows us to avoid prior segmentation. As a result, our algorithm estimates accurate depth maps with high throughput.  
\end{itemize}

This paper is organized as follows. In Section \ref{sec:relatedworks}, we describe similar techniques that estimate depth maps using consecutive images for dynamic scenes and highlight why they are insufficient for our purpose. We then provide an overview of our depth map estimation algorithm in Section \ref{sec:method}. In Section \ref{sec:motionestimation}, we detail how we use the prior depth information to estimate independent rigid motions and describe how we assign them to the pixels of the previous depth map in Section \ref{sec:motionassign}. In Section \ref{sec:results}, we evaluate the performance of our approach using commonly used datasets. This is followed by Section \ref{sec:discussion}, where we compare our approach to similar techniques and describe our limitations. Finally, we summarize our key contributions and conclude our paper in Section \ref{sec:conclusion}.

\begin{table*}[ht]
   \centering
   \begin{tabular}{lcc|cc|ccc}
   \toprule
   \multicolumn{1}{c}{} & \multicolumn{1}{c}{} & \multicolumn{1}{c|}{} & \multicolumn{2}{c|}{\underline{\textbf{High Latency}}} & \multicolumn{3}{c}{\underline{\textbf{High Computational Complexity}}}  \\
   \multicolumn{1}{c}{\textbf{Method}}                           & \multicolumn{1}{c}{\textbf{Category}}  & {\textbf{Prior Depth?}}   &\textbf{Non-Causal}  & \textbf{Multi-Frame} & \textbf{Dense Optical Flow} & \textbf{Segmentation} & \textbf{Deep Neural Network} \\ \midrule
   \textbf{This Work}                                 &                    &$\checkmark$           &                    &&                        &                       &                                                                       \\ \hline 
   Choi \emph{et al.} \cite{Choi2010}        & DTM                 &$\checkmark$           &$\checkmark$            &$\checkmark$                     & $\checkmark$                    &                       &                                           \\ \hline
   Wang \emph{et al.} \cite{Wang2010}        & DTM                 &$\checkmark$           &        &                   & $\checkmark$                    &                       &                                           \\ \hline
   Li \emph{et al.} \cite{Li2011}            & DTM                 &$\checkmark$           &        &                    & $\checkmark$                    &                       &                                           \\ \hline
   Karsch \emph{et al.} \cite{Karsch2014}    & DTM                 &$\checkmark$           &        &           & $\checkmark$                    &                       &                                   \\ \hline
   Zhang \emph{et al.} \cite{Zhang2014}      & DTM                 &$\checkmark$           &$\checkmark$            &$\checkmark$                    & $\checkmark$                    &                       &                                           \\ \hline 
   Zhang \emph{et al.} \cite{Zhang2011}      & NRSFM              &                       &        &$\checkmark$                    &                             &$\checkmark$                       &                                   \\ \hline
   Roussos \emph{et al.} \cite{Roussos2012}  & NRSFM              &                       &        &$\checkmark$                    & $\checkmark$                    & $\checkmark$              &                                           \\ \hline
   Ranftl \emph{et al.} \cite{Ranftl2016}    & NRSFM              &                       &        &           & $\checkmark$                    & $\checkmark$              &                                           \\ \hline
   Kumar \emph{et al.} \cite{Kumar2017}      & NRSFM              &                       &        &           & $\checkmark$                    &                       &                                   \\ \hline
   Kumar \emph{et al.} \cite{Kumar2019}      & NRSFM              &                       &        &           & $\checkmark$                    &                       &                                   \\ \hline
   Casser \emph{et al.} \cite{Casser2018}    & NNFDE              &                       &        &$\checkmark$                    &                             & $\checkmark$              &                     $\checkmark$              \\ \hline
   Gordon \emph{et al.} \cite{Gordon2019}    & NNFDE              &                       &        &           &                             &                       &               $\checkmark$              \\ \hline
   Li \emph{et al.} \cite{Li2019}            & NNFDE              &                       &         &           & $\checkmark$                    & $\checkmark$              &                     $\checkmark$              \\ \bottomrule
   \end{tabular}
   \caption{\textbf{Related Works Comparison}: We summarize the techniques in Section \ref{sec:relatedworks} that estimate dense depth maps for dynamic scenes. For each technique, we use a checkmark to highlight features that are related to its latency and computational complexity. As our goal is to estimate depth maps with low latency and high throughput, we see that these previous techniques are insufficient.}\label{tab:methodtables}
\end{table*}

\section{Related Work}\label{sec:relatedworks}
Many approaches have been proposed to estimate depth in dynamic scenes for a variety of different applications. Given its breadth, we only summarize techniques that estimate depth with a similar setup or those that only use consecutive and monocular images. As stated in the previous section, many of these approaches are insufficient for the applications that we consider because they estimate depth maps with either high latency or high computational complexity. We list these methods in Table \ref{tab:methodtables}, and we highlight the different features that pertain to latency and computational complexity. For latency, we note whether the technique is non-causal or if it is a multi-frame approach (requiring more than 2 consecutive frames to estimate a depth map). For computational complexity, we note whether the technique requires dense optical flow, segmentation, or is a deep neural network based approach. These features are all computationally expensive \cite{Fortun2015,Yao2019}.

\subsection{Depth Transfer Methods (DTM)}
Similar to our approach, the depth transfer methods estimate new depth maps using previously measured ones \cite{Wang2010,Choi2010,Li2011,Zhang2014,Karsch2014}. However, instead of estimating the 3D motion in the scene, these approaches instead estimate the \emph{dense} optical flow between the current image and one that corresponds to a previous depth map and use the optical flow to warp the previous depth map to obtain a new one. 

The authors of \cite{Wang2010,Choi2010,Li2011,Zhang2014} use this framework to equalize the frame rate between recorded image and depth video. They estimate new depth maps using frames where both images and depth maps are available, and these techniques have the advantage that they have data from both the preceding and \emph{future} frames. While these techniques are effective for increasing the frame rates of depth videos, they are not causal. While Wang \emph{et al.} \cite{Wang2010} and Li \emph{et al.} \cite{Li2011} can be adapted to causally estimate depth using only two frames, these approaches do not account for changes in depth since the preceding depth maps are simply warped. This is sufficient for small changes in depth or in-plane motion, but it cannot be generalized to all dynamic scenes, which we demonstrate in Section \ref{sec:discussion}. 

Karsch \emph{et al.} \cite{Karsch2014} similarly warps a previously measured depth map, but differs in that it uses depth maps taken from a training set of image and depth map pairs of similar scenes. One benefit of this approach is that it can support monocular depth estimation. To estimate the depth maps for consecutive images, these authors improve accuracy by using motion cues to estimate temporally consistent depth maps. This method fails when the training set does not contain image and depth map pairs of similar scenes. As our approach does not rely on a training set, it does not suffer from this issue.

\subsection{Non-Rigid Structure-from-Motion (NRSFM)}

Non-rigid structure-from-motion techniques estimate relative depth using only images by exploiting statistical and physical heuristics\cite{Jensen2018}. Here, we focus on the methods that estimate dense depth maps in dynamic scenes, namely \cite{Zhang2011,Roussos2012,Ranftl2016,Kumar2017,Kumar2019}. Like our approach, these methods are all causal and with the exception of \cite{Zhang2011,Roussos2012} estimate depth using only two consecutive frames. 

However, unlike our approach, these techniques have high complexity. The authors of \cite{Zhang2011,Roussos2012,Ranftl2016} estimate depth by first segmenting the pixels into rigid regions. Zhang \emph{et al.} \cite{Zhang2011} directly segments the pixels whereas Roussos \emph{et al.} \cite{Roussos2012} and Ranftl \emph{et al.} \cite{Ranftl2016} first compute a dense optical flow field and then segment it to initialize their algorithms. This comes with a high computational cost and lowers the throughput at which depth can be estimated, a detriment for many applications. However, this rigid motion segmentation is necessary because estimating depth is underdetermined without geometric assumptions like rigidity. Noting the challenge of rigid motion segmentation, other approaches use simpler partitions \cite{Kumar2017,Kumar2019}. Kumar \emph{et al.} \cite{Kumar2017} assumes that dynamic scenes are locally rigid and partitions the scene into rigid superpixels. However, this approach still requires dense optical flow to estimate depth. Our approach is similar to these techniques in that it models the dynamic scene using rigid motions, but avoids the need of prior segmentation or partitioning because it leverages the previous depth map to both estimate the rigid motions and assign them to the pixels of the previous depth map. As we will show, this enables efficient depth map estimation.

\subsection{Neural Networks for Depth Estimation (NNFDE)}
In addition to the previous approaches, many deep neural networks have recently been proposed to estimate depth using monocular images. We focus on techniques that use consecutive images to estimate depth in dynamic scenes \cite{Casser2018,Gordon2019,Li2019}. In addition to consecutive images, some of these methods require prior segmentation as an input before depth can be estimated. Casser \emph{et al.} \cite{Casser2018} and Li \emph{et al.} \cite{Li2019} require object-level segmentation masks whereas Gordon \emph{et al.} require bounding boxes of possibly mobile objects. The technique in \cite{Li2019} even requires an initial depth map obtained using structure-from-motion techniques, underscoring the inherent difficulty of estimating depth for dynamic scenes. While these techniques are promising, they are computationally complex, and their performance is limited by the diversity of their training set. Unlike these approaches, our technique estimates depth by exploiting physical heuristics and does not require a training set.

Across a variety of approaches that estimate depth in dynamic scenes, we see that many methods require a dense optical flow field or prior segmentation. As summarized in Table \ref{tab:methodtables}, none of these approaches satisfy our requirements.  In the next sections, we show how we can avoid these operations for our problem setup to efficiently estimate depth in dynamic scenes.

\begin{figure*}[t]
   \centering
   \includegraphics{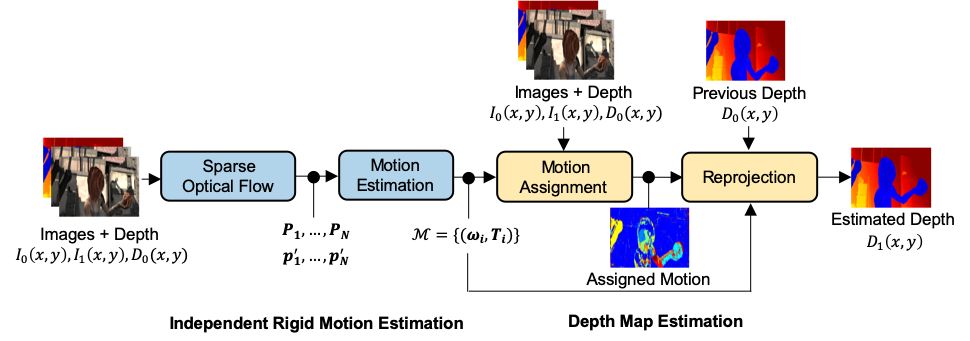}
   \caption{\textbf{Algorithm Pipeline}: Our algorithm takes as input two consecutive images and a previous depth map. It uses these inputs to first estimate the independent and rigid motions in the scene using the sparse optical flow computed from the images and its corresponding depth. Our technique then obtains a depth map by assigning the estimated rigid motions to the appropriate pixels of the previous depth map, guided by the photometric error obtained by reprojecting the images.}
   \label{fig:pipeline}
\end{figure*}

\section{Depth Map Estimation Overview}\label{sec:method}

As shown in Figure \ref{fig:setup}, our algorithm takes as input two consecutive images and a previous depth map. Our goal is to output a new depth map that corresponds to the current image. The pipeline of our algorithm is depicted in Figure \ref{fig:pipeline} and is composed of two major stages: one that estimates the rigid motions in the scene (Section \ref{sec:motionestimation}) and another that assigns them to obtain a new depth map (Section \ref{sec:motionassign}).

\subsection{Notation} 

We denote the image and depth map in the previous frame as $I_0(x,y)$ and $D_0(x,y)$, respectively, and the current image as $I_1(x,y)$. Our goal is to estimate the current depth map, denoted as $D_1(x,y)$. The 2D coordinate of the $i^{\text{th}}$ pixel in the previous frame is denoted as $\boldsymbol{p_i}=(x_i,y_i)^T$ and its correspondence in the current frame is $\boldsymbol{p'_i}=(x'_i,y'_i)^T$. To distinguish between vectors and scalars, we will bold the former. 

We assume that our camera is calibrated and images are formed by perspectivity. This allows us to relate the components of $\boldsymbol{p_i}$ to its 3D coordinate, $\boldsymbol{P_i}=(X_i,Y_i,Z_i)^T$, as follows:
\begin{equation}\label{eq:perspective}
   \frac{X_i}{Z_i} = \frac{x_i-x_c}{f} \qquad \frac{Y_i}{Z_i} = \frac{y_i-y_c}{f}
\end{equation} 
where $f$ is the principal distance (or focal length) and $(x_c,y_c)$ is the principal point. We assume that the previous image and depth map are spatially aligned so that $Z_i=D_0(x_i,y_i)$ in \eqref{eq:perspective}. This means that we have the 3D coordinate for each pixel in the previous frame.

From the relationship in \eqref{eq:perspective}, we define the projection operator, $\pi\left(\boldsymbol{P_i}\right)=\boldsymbol{p_i}$, to map a 3D point to its pixel coordinate. We also make use of standard linear algebra notation. We denote $\boldsymbol{\hat{x}}, \boldsymbol{\hat{y}}, \boldsymbol{\hat{z}}$ as the unit vectors oriented along the 3D coordinate axes. We also use standard operations like the dot product, e.g. $\boldsymbol{P_i}\cdot\boldsymbol{P_j}$, and the cross product, $\boldsymbol{P_i}\times\boldsymbol{P_j}$.

\begin{figure*}[t]
   \centering
   \includegraphics{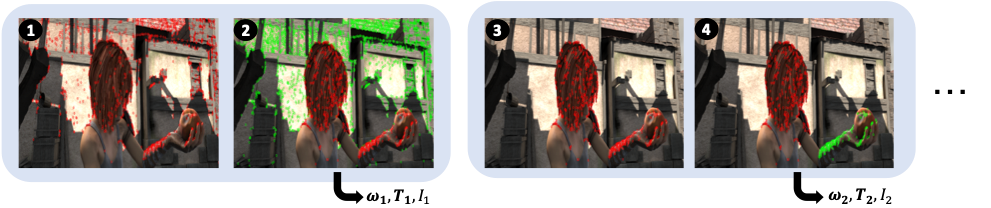}
   \caption{\textbf{Sequential Rigid Motion Estimation}: We depict how the estimated rigid motions are obtained. In the first image, we highlight the pixels (shown in red) where $\boldsymbol{P_i}$ and $\boldsymbol{p_i}'$ are known. We use RANSAC to estimate the rigid motion with the largest inlier set, shown in green in the second image, and remove these pixels from further consideration. This process is repeated, as shown in the third and fourth images, until the size of the inlier set falls below a threshold.} 
   \label{fig:motion_est}
\end{figure*}

\section{Independent Rigid Motion Estimation}\label{sec:motionestimation}

As we previously stated, our technique assumes that the motion in dynamic scenes can be approximated using independent rigid motions. In order to estimate it, we invert an image formation model that relates the 3D motion in the scene to its 2D displacement across the consecutive images. In this section, we describe how we accomplish this.

\subsection{Sparse Optical Flow}
Unlike the approaches described in Section \ref{sec:relatedworks}, our technique estimates the 3D scene motion using the optical flow at a sparse set of pixels. To do so, we first detect corners in the previous image by using the FAST corner detector \cite{Rosten2005} and then estimate the optical flow at these pixels using the Lucas Kanade algorithm \cite{lucaskanade}. We use these algorithms because they are computationally efficient. 

If the optical flow for the $i^{\text{th}}$ pixel in the previous frame, or $\boldsymbol{p_i}$, is known, then its correspondence in the current frame, $\boldsymbol{p'_i}$, can be trivially found. Furthermore, for every pixel detected by the FAST corner detector, we also compute its 3D coordinate, $\boldsymbol{P_i}$, by using the previous depth map, $D_0(x,y)$, as described in Section \ref{sec:method}.  

\subsection{Motion Estimation} 
With the optical flow estimated, we now describe how the independent rigid motions are obtained. In the simplest case, there is a single rigid motion, and we first describe how to estimate it. We then extend our approach to handle multiple independent rigid motions.

\subsubsection{Single Rigid Motion}
Our approach uses a linear representation of rigid motion, namely angular and translational velocity, denoted as $\boldsymbol{\omega}, \boldsymbol{T}\in\mathcal{R}^3$, respectively. Our goal is to relate these parameters to the pixel-wise displacements across the consecutive frames. We can obtain this expression by first applying the rigid motion to $\boldsymbol{P_i}$, the 3D coordinate of the $i^{\text{th}}$ pixel which we previously computed, to obtain $\boldsymbol{P'_i}$, its 3D position in the current frame, where: 
\begin{equation}\label{eq:rigidmotion}
	\boldsymbol{P_i}'= \boldsymbol{P_i} + \boldsymbol{\omega}\times\boldsymbol{P_i} + \boldsymbol{T}
\end{equation}  
We then equate its projection to its pixel location in the current frame, or $\boldsymbol{p_i}'=(x'_i,y'_i)$, using \eqref{eq:perspective}. By manipulating the terms, we arrive at the following relationships:    
\begin{equation}\label{eq:phix}
	\underbrace{(x'_i-x_c)\boldsymbol{\hat{z}}\cdot\boldsymbol{P'_i}-f\boldsymbol{\hat{x}}\cdot\boldsymbol{P'_i}}_{\phi_x(\boldsymbol{P_i},\boldsymbol{p_i}',\boldsymbol{\omega},\boldsymbol{T})}= 0
\end{equation}
\begin{equation}\label{eq:phiy}
	\underbrace{(y'_i-y_c)\boldsymbol{\hat{z}}\cdot\boldsymbol{P'_i}-f\boldsymbol{\hat{y}}\cdot\boldsymbol{P'_i}}_{\phi_y(\boldsymbol{P_i},\boldsymbol{p_i}',\boldsymbol{\omega},\boldsymbol{T})} = 0
\end{equation}
These expressions \emph{linearly} relate the rigid motion parameters ($\boldsymbol{\omega},\boldsymbol{T}$), which we want to estimate, to the 3D coordinate of the $i^{\text{th}}$ pixel in the previous frame ($\boldsymbol{P_i}$) and its correspondence in the current frame ($\boldsymbol{p_i}'$). We can obtain these parameters by minimizing \eqref{eq:optimize} in a least squares sense:
\begin{equation}\label{eq:optimize}
   \boldsymbol{\omega}^*,\boldsymbol{T}^* = \underset{\boldsymbol{\omega},\boldsymbol{T}}{\mathrm{argmin}}~\sum_{i=1}^N\phi_x^2(\boldsymbol{P_i},\boldsymbol{p_i}',\boldsymbol{\omega},\boldsymbol{T})+\phi_y^2(\boldsymbol{P_i},\boldsymbol{p_i}',\boldsymbol{\omega},\boldsymbol{T})
\end{equation}
where $N$ is the total number of pixels (in our case, those detected by FAST). The solution to \eqref{eq:optimize} can be found efficiently and is equivalent to solving a $6\times6$ linear system. Unlike standard approaches (e.g., eight point algorithm), we estimate the rigid motion parameters in absolute units and only need 3 sets of correspondences to obtain it. In practice, because of potential errors in the optical flow estimates, we use RANSAC \cite{Fischler1981} to robustly estimate the rigid motion. By lowering the number of correspondences required to obtain a rigid motion hypothesis, we minimize the likelihood of selecting erroneous correspondences and enable RANSAC to robustly estimate the rigid motion.   

\subsubsection{Estimating Multiple Rigid Motions} 

To estimate the multiple and independent rigid motions in the scene, we sequentially estimate them individually using RANSAC as previously described. In addition to obtaining the rigid motion, RANSAC also determines the inlier set, which is defined as:
\begin{equation}\label{eq:reprojerr}
	\mathcal{I} = \left\{ i :  \left|\left|\pi\left(\boldsymbol{P'_i}\right) - \boldsymbol{p_i}' \right|\right|_2^2 \leq \epsilon\right\}
\end{equation}
where $\epsilon$ is a threshold based on the projection error. 

In our approach, we adapt RANSAC to estimate $\boldsymbol{\omega}$ and $\boldsymbol{T}$ that \emph{maximizes} the size of the inlier set. We then remove the pixels in the inlier set from further consideration and repeat this process to greedily estimate the rigid motions as shown in Figure \ref{fig:motion_est}. This is done to increase the diversity of the estimated rigid motions to best represent the dynamic motion in the scene and to reduce the complexity of the motion assignment process, described in the next section. This process is repeated until the size of the inlier set falls below a minimum size, $N_{\text{min}}$. As a result, our approach does not need the number of rigid motions to be specified. The output of this algorithm is the set of estimated rigid motions, which we denote as $\mathcal{M} = \{(\boldsymbol{\omega_i},\boldsymbol{T_i})\}$. We summarize our approach in Algorithm \ref{alg:rm_cluster}.  

\begin{algorithm}[t]
\caption{Independent Rigid Motion Estimation}\label{alg:rm_cluster}
\begin{algorithmic}[1]
\Require $I_0(x,y)$, $I_1(x,y)$, and $D_0(x,y)$; $N_{\text{min}}$, $\epsilon$
\Ensure $\mathcal{M} = \{(\boldsymbol{\omega_i},\boldsymbol{T_i})\}$

\State $\mathcal{P} \gets \{(x_i,y_i)\}$ using FAST corner detector on $I_0$
\State $\mathcal{P}' \gets \{(x'_i,y'_i)\}$ using Lucas Kanade on $\mathcal{P}, I_0$ and $I_1$  
\State $\mathcal{Z}\gets \{D_0(x_i,y_i) \text{ for } (x_i,y_i) \in \mathcal{P}\}$
\State  $\mathcal{M} \gets \{\empty\}$

\Repeat 
	\State $\boldsymbol{\omega_i},\boldsymbol{T_i},\mathcal{I}_i \gets $ Estimate rigid motion using RANSAC
	\If{$|\mathcal{I}_i| > N_{\text{min}}$}
		\State Remove inlier values from $\mathcal{P}, \mathcal{P}', \mathcal{Z}$
		\State $\mathcal{M} \gets \mathcal{M} \cup (\boldsymbol{\omega_i},\boldsymbol{T_i})$ 
	\EndIf
\Until $|\mathcal{I}_i| < N_{min}$ or $\mathcal{P}$ is empty
\end{algorithmic}
\end{algorithm}

\section{Depth Map Estimation}\label{sec:motionassign}

Once the rigid motions are estimated, we obtain a new depth map by using these estimated rigid motions to reproject the previous depth map. To do so, we obtain the 3D position of each pixel, apply the appropriate rigid motion, and project its updated depth. In this section, we describe how we determine which estimated rigid motion to use in order to obtain a new depth map. 

\subsection{Motion Assignment}

\begin{figure*}[t]
   \centering
   \includegraphics{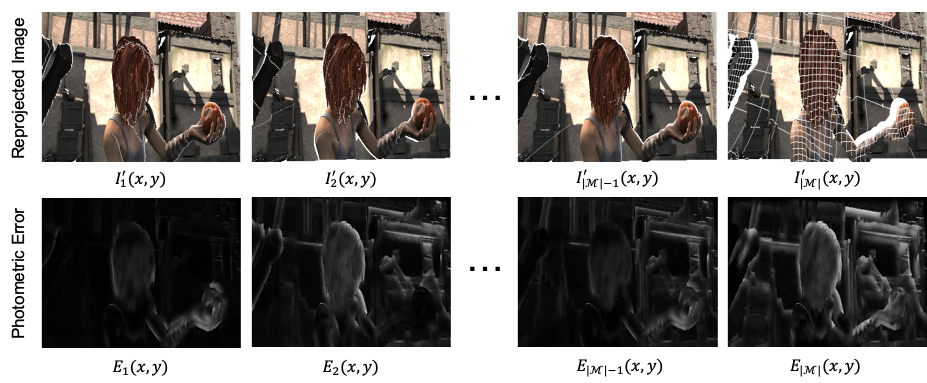}
   \caption{\textbf{Photometric Error}: We reproject the previous image using the previous depth map and the estimated rigid motions as shown in the first row. The photometric error, shown in the second row, is then obtained by computing the absolute difference between the reprojected images and the current one and filtering this output using a guided filter.}
   \label{fig:photometric_example}
\end{figure*}

To determine the motion assignment, we exploit the fact that in the previous frame, both the image and the depth map are spatially aligned. This allows us to reproject the previous image. This is important because if a set of pixels move with a certain rigid motion, then its reprojection must coincide with its corresponding pixels in the next frame. Consequently, the pixel-wise difference between these reprojected pixels and its correspondences in the next frame, or the photometric error, must have a low magnitude. Our approach uses this insight to assign the estimated rigid motions to the appropriate pixels of the previous depth map. 

We begin by reprojecting the previous image, or $I_0(x,y)$, using each of the estimated rigid motions. We obtain the 3D coordinate of each pixel in the previous image, or $\boldsymbol{P}_i$, by using \eqref{eq:perspective} and $D_0(x,y)$. Given the $j^{\text{th}}$ estimated rigid motion, we first compute $\boldsymbol{P}'_i$ using \eqref{eq:rigidmotion}, from which the reprojected image, denoted as $I'_j(x,y)$, can be defined as follows:
\begin{equation}\label{eq:reprojected}
   I'_j(x'_i,y'_i) = I_0(x_i,y_i)
\end{equation}
where $(x'_i,y'_i)^T=\pi\left(\boldsymbol{P}'_i\right)$. The photometric error is the absolute difference between this reprojected image and the current one, $I_1(x,y)$, and we define it as:
\begin{equation}\label{eq:photometric}
	E_j(x_i,y_i) = \left|I_1(x'_i,y'_i)-I'_j(x'_i,y'_i)\right|
\end{equation} 
with $(x'_i,y'_i)$ similarly defined. To ensure that the photometric error is locally smooth, we also filter $E_j(x,y)$ with a guided filter \cite{He2013} and use $I_0(x,y)$ as the guide image. We show examples of the reprojected image and the resulting photometric error in Figure \ref{fig:photometric_example}.
 
\begin{figure}[t]
   \centering
   \includegraphics{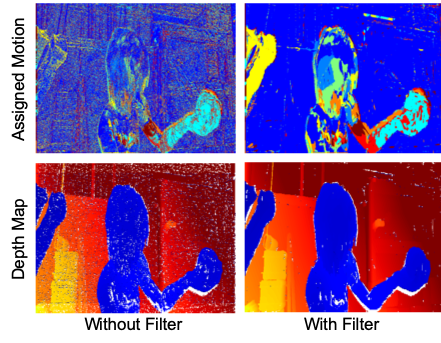}
   \caption{\textbf{Motion Assignment}: We show the motion assignment, where different colors represent the different estimated rigid motions, and the resulting depth map. Without filtering, the estimated rigid motions are spuriously assigned and results in artifacts in the depth map. When the photometric error is filtered, we see that the motion assignment is locally smooth and the depth map contains less artifacts.}
   \label{fig:label_assignment}
\end{figure}

Finally, we assign the $j^\text{th}$ estimated rigid motion to the $i^{\text{th}}$ pixel if the following holds:
\begin{equation}\label{eq:poselabel}
	j = \underset{1 \leq k \leq |\mathcal{M}|}{\mathrm{argmin}}~E_k(x_i,y_i)
\end{equation}
We solve \eqref{eq:poselabel} for every pixel in the previous depth map, and we visualize an example of this motion assignment in Figure \ref{fig:label_assignment}. This figure also shows the impact of the guided filtering. In our experiments, we find that filtering the photometric error is essential and helps ensure that the motion assignment is also locally smooth, which agrees with our intuition that dynamic scenes are locally rigid. Furthermore, we also see that the resulting depth maps also have less artifacts, and  we discuss the cause of these artifacts in Section \ref{sec:limitations}. Our approach is similar to the framework proposed in \cite{Rhemann2011}, which applied the same filtering operations for optical flow estimation and stereo matching. 

In the processing of assigning the estimated rigid motions, our algorithm also segments the rigid motion in the scene. This is in contrast to some of the approaches in Section \ref{sec:relatedworks}, which require dense optical flow estimation and rigid motion segmentation before depth can be estimated. We are able to do this without these operations because we have a previous depth map, which allows us to compute the photometric error to determine the best rigid motion assignment. Furthermore, this process is also computationally efficient. Computing and filtering the photometric error in \eqref{eq:photometric} has $O(n)$ complexity, where $n$ is the total number of pixels in the previous depth map. This is repeated for each of the $|\mathcal{M}|$ estimated rigid motions, and this computation can be parallelized. We note that $|\mathcal{M}| \leq \frac{N}{N_{\text{min}}}$, where $N$ and $N_{\text{min}}$ are the parameters used to estimate the rigid motions in Section \ref{sec:motionestimation}. In our experiments, we choose these parameters to balance the accuracy of the estimated depth maps with its throughput. As a result of this, we are able to estimate dense depth maps in real-time, or under 33.3 milliseconds for each depth map, on a standard laptop computer (2.7 GHz i5-5257U cores) compared to the \emph{minutes} reported by the methods summarized in Section \ref{sec:relatedworks}.    

\subsection{Reprojection}
Finally, once the estimated rigid motions are assigned, we estimate the current depth map by reprojecting the previous one. For every pixel in the previous depth map, we first compute its 3D coordinate, $\boldsymbol{P_i}$, using \eqref{eq:perspective} and then compute $\boldsymbol{P_i}'$ using its rigid motion determined by \eqref{eq:poselabel}. The depth map is finally obtained as follows:
\begin{equation}
   D_1(x',y') = \boldsymbol{\hat{z}}\cdot\boldsymbol{P_i}'
\end{equation}
where $(x',y')^T=\pi\left(\boldsymbol{P_i}'\right)$. When multiple 3D points are reprojected to the same pixel location, we retain the smaller depth value. We summarize our depth estimation process  in Algorithm \ref{alg:assignment}. 

\begin{algorithm}[t]
\caption{Depth Map Estimation}\label{alg:assignment}
\begin{algorithmic}[1]
\Require $I_0(x,y),I_1(x,y),$ and $D_0(x,y)$; $\mathcal{M}=\{(\omega_i,t_i)\}$ 
\Ensure $D_1(x,y)$

\State $ j\gets 1$
\Repeat
   \State Reproject $I_0(x,y)$ using $\boldsymbol{\omega_j}, \boldsymbol{T_j},$ and $D_0(x,y)$ 
	\State Compute and filter $E_j(x,y)$ using \eqref{eq:photometric}
	\State $j \gets j+1$
\Until $ j = |\mathcal{M}|$
\Repeat  
	\State Compute the best motion $j$ using \eqref{eq:poselabel}
	\State Compute $\boldsymbol{P'_i}$ from $\boldsymbol{P_i}$, $\boldsymbol{\omega_{j}}$ and $\boldsymbol{T_{j}}$ using \eqref{eq:rigidmotion}
   \State $(x'_i,y'_i)^T\gets\pi(\boldsymbol{P'_i})$
   \State $D_1(x'_i,y'_i) \gets \boldsymbol{\hat{z}}\cdot \boldsymbol{P'_i}$
\Until all pixels are reprojected
\end{algorithmic}
\end{algorithm}

\section{Algorithm Evaluation}\label{sec:results}
To evaluate our algorithm, we use RGB-D datasets that contain calibrated image and dense depth map pairs of different dynamic scenes. These datasets are also used to evaluate the approaches in Section \ref{sec:relatedworks} and include:
\begin{itemize}
   \item \textbf{Deformable Surfaces (DS)} \cite{Varol2009}: This dataset contains real sequences of objects undergoing non-rigid deformations. We use the \emph{kinect\_paper} and \emph{kinect\_tshirt} sequences.
   \item \textbf{MPI Sintel (MPI)} \cite{Butler:ECCV:2012}: This dataset contains synthetic scenes with both articulated and camera motion. We use the clean video sequences of \emph{alley\_1}, \emph{ambush\_7}, \emph{bandage\_1}, \emph{bandage\_2}, \emph{shaman\_2}, \emph{shaman\_3}, \emph{sleeping\_1}, and \emph{sleeping\_2}.    
   \item \textbf{TU Munich RGB-D (TUM)} \cite{Sturm2012}: This dataset is typically used to benchmark SLAM algorithms. We use the sequences in the \emph{Dynamic Objects} category, which contain both camera and human motion.
   \item \textbf{Virtual KITTI (VKITTI)} \cite{Gaidon2016}: This dataset contains synthetic scenes from the perspective of a car driving through different urban environments. We use the overcast sequences for \emph{1}, \emph{2}, \emph{6}, \emph{18}, and \emph{20}.   
\end{itemize}

\begin{figure*}[ht]
   \centering
   \begin{subfigure}{.32\textwidth}
      \centering
      \includegraphics[scale=.65]{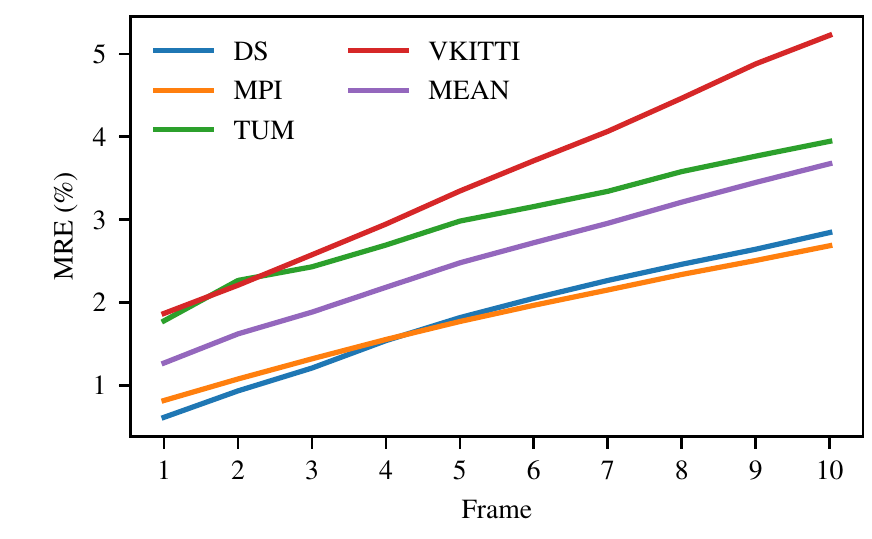}
   \end{subfigure} 
   \begin{subfigure}{.32\textwidth}
      \centering
      \includegraphics[scale=.65]{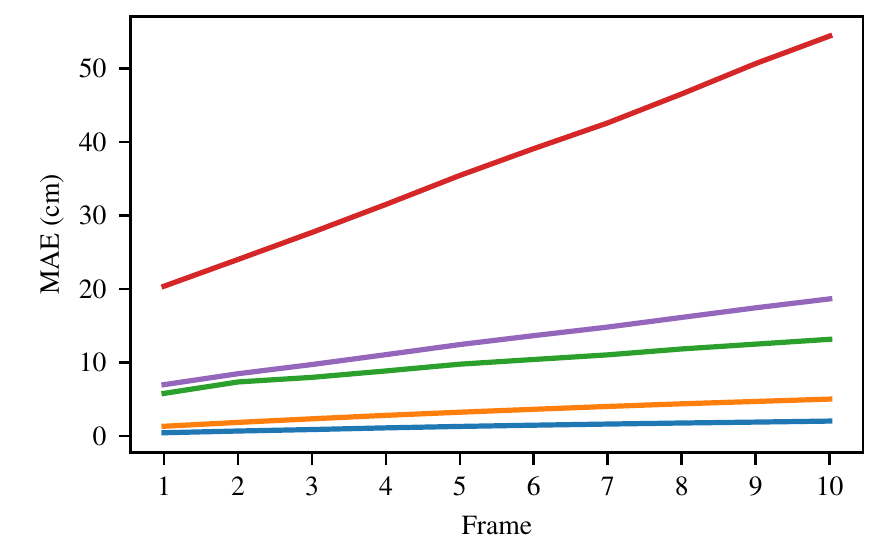}
   \end{subfigure} 
   \begin{subfigure}{.32\textwidth}
      \centering
      \includegraphics[scale=.65]{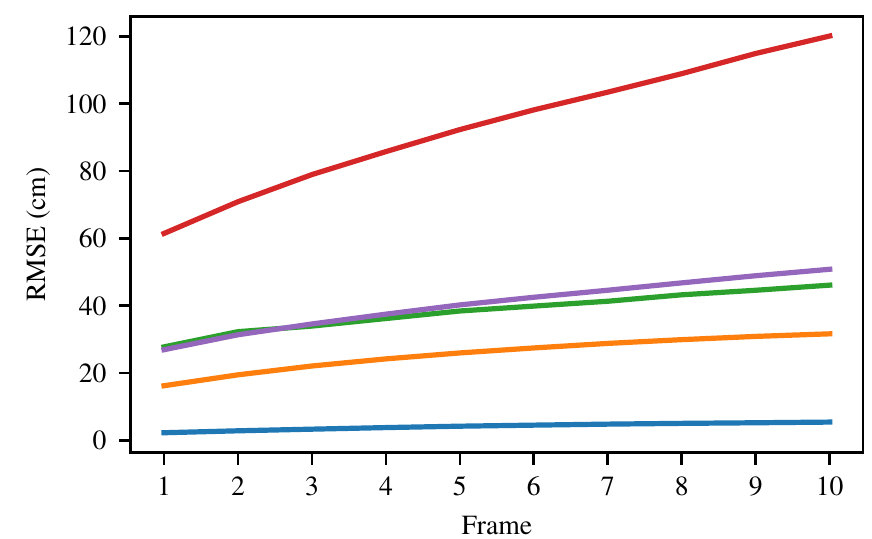}
   \end{subfigure}
   \caption{\textbf{Sequential Estimation Results}: The average MRE, MAE, and RMSE of consecutively estimated depth maps are plotted for the datasets we evaluate on.} 
   \label{fig:multiframe_results}
\end{figure*}

\begin{figure*}[ht]
   \centering
   \includegraphics{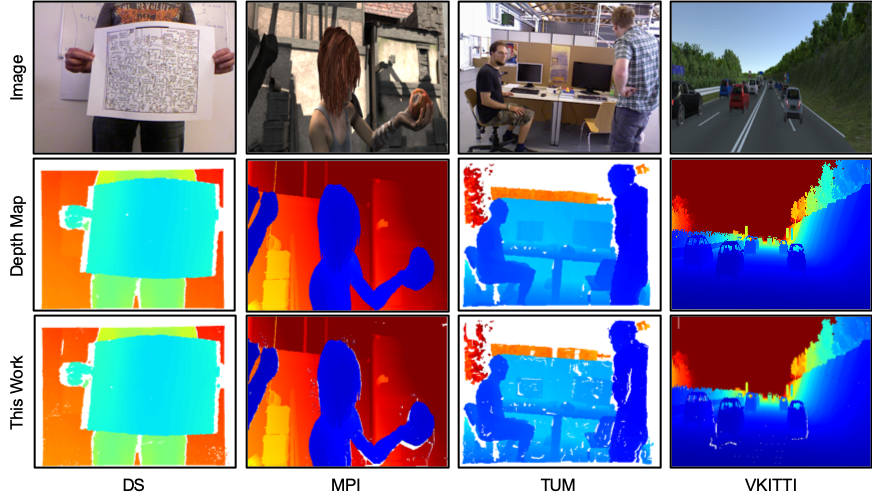}
   \caption{\textbf{Estimated Depth Maps}: We compare our estimated depth maps (This Work) against the ground truth. The white areas indicate regions where depth is unavailable. We discuss the cause of these regions in our estimated depth maps in Section \ref{sec:limitations}. }
   \label{fig:examples}
\end{figure*}

\subsection{Methodology}
We test our approach by estimating depth maps sequentially: we first estimate a new depth map using its corresponding image and a previous image and measured depth map pair, and for consecutive frames, we then use the \emph{estimated} depth map to obtain the next one. 

To evaluate the estimated depth maps, we compute the mean relative error (MRE). This metric penalizes a unit error at a close range more than that further away, which is an appropriate metric for applications that use depth to interact with its immediate environment. The MRE also allows us to compare the performance of our algorithm across datasets that have different dynamic ranges. To highlight the different dynamic ranges for the different datasets, we also compute the mean absolute error (MAE) and the root mean squared error (RMSE). These metrics are defined as follows: 
\begin{equation}\label{eq:mre}
    \text{MRE} = \frac{1}{N}\sum_{i=1}^N \frac{|Z_i-\hat{Z}_i|}{Z_i} 
\end{equation}
\begin{equation}\label{eq:mae}
    \text{MAE} = \frac{1}{N}\sum_{i=1}^N |Z_i-\hat{Z}_i|
\end{equation}
\begin{equation}\label{eq:mae}
    \text{RMSE} = \sqrt{\frac{1}{N}\sum_{i=1}^N (Z_i-\hat{Z}_i)^2}
\end{equation}
where $\hat{Z}_i$ and $Z_i$ are the estimated and measured depth for the $i^{\text{th}}$ pixel, respectively, and $N$ is the total number of depth estimates. In our evaluation, we compute these metrics for pixels with ground truth depth values that are within 20 meters. We compute these metrics for 10 sequentially estimated depth maps and average them over 100 different starting points for the sequences in each dataset (except the MPI dataset, where each sequence only has 50 frames). 

\subsection{Implementation Details}
\begin{table}
   \centering
   \begin{tabular}{lccc} \hline
      \toprule
      \textbf{Dataset} & \textbf{Resolution} & \textbf{Frame Rate (FPS)} & \textbf{Time Per Frame (ms)} \\
      \midrule
      {DS} & $640\times480$	&	32	 & 31.3\\	
      {MPI} & $1024\times436$	&	12	 & 83.3\\
      {TUM} & 	$640\times480$	& 34	& 29.4\\
      {VKITTI} & 	$1242\times375$ &	14	 & 71.4\\
      \bottomrule	
   \end{tabular}
   \caption{\textbf{Throughput}: We summarize the median frame rate and the estimation time per frame for the sequences of each dataset as profiled on a standard laptop computer (2.7 GHz i5-5257U cores). This is significantly faster than previous approaches, like Kumar \emph{et al.} \cite{Kumar2017}, which require \emph{several minutes} to estimate a $1024\times436$ depth map.}
   \label{tab:throughput}
\end{table}

We implement our algorithm following the details stated in Sections \ref{sec:motionestimation} and \ref{sec:motionassign} and tune algorithm parameters separately for each dataset. Whenever possible, we use OpenCV to implement our approach. As shown in Table \ref{tab:throughput}, our code estimates dense depth maps in real-time, or over $30$ frames per second (FPS), for the DS and TUM datasets and in near real-time for the other datasets on a standard laptop computer (2.7 GHz i5-5257U cores). This is significantly faster than the approaches in Section \ref{sec:relatedworks}, which report \emph{several minutes} to estimate depth maps of the same resolution. Our computation is dominated by the motion assignment, where approximately 75\% of the time is spent on assigning the estimated rigid motions, and the remaining 25\% of the time is spent on estimating the rigid motions.

\subsection{Results}
\begin{table}[t]
   \centering
   \begin{tabular}{lccc} \hline
      \toprule
      \textbf{Dataset} & \textbf{MRE (\%)} & \textbf{MAE (cm)} & \textbf{RMSE (cm)} \\
      \midrule
      {DS} & 1.8	&	1.3	&	4.1 \\	
      {MPI} & 1.8	&	3.3	&	25.6 \\
      {TUM} & 	3.0	& 9.9	&	38.4 \\
      {VKITTI} & 	3.5	&	37.2	&	93.5 \\
      \midrule
      \textbf{Mean} & 2.5 & 12.9 & 40.4 \\
      \bottomrule	
   \end{tabular}
   \caption{\textbf{Results}: We summarize the performance of our approach on each dataset by averaging the different metrics over the frames we estimate.}
   \label{tab:results}
\end{table}

The results of our evaluation are shown in Figure \ref{fig:multiframe_results}, where we plot the MRE, MAE, and RMSE for the depth maps as they are sequentially estimated. Additionally, we also average the MRE, MAE, and RMSE over the frames for each dataset in Table \ref{tab:results} and show examples of the estimated depth maps in Figure \ref{fig:examples}. 

We observe that all of the error metrics increase as more consecutive depth maps are estimated, and this is due to the errors in the motion estimation and assignment that accumulate across frames. This is especially pronounced in the TUM and VKITTI datasets, which have higher errors than the DS and MPI datasets. For the TUM dataset, the images are affected by motion blur and are not perfectly time synchronized with the depth maps. This impacts the motion assignment as described in Section \ref{sec:motionassign}. For the VKITTI dataset, the difference between the foreground and background depth is large. Therefore, errors at these boundaries between the foreground and background (due to erroneous motion assignments, for example) are high.     

Nonetheless, our algorithm is still accurate, and when averaged across the different datasets, we see that our algorithm obtains a MRE between 1.3\%-3.7\%. Moreover, our results also suggest that we can reduce the usage of the depth sensor by over 90\% but still estimate depth maps within 2.5\% of the ground truth for general scenes.

\section{Discussion}\label{sec:discussion}
In this section, we compare our method to previous approaches (Section \ref{sec:baselinecomparison}) and describe our limitations (Section \ref{sec:limitations}).

\subsection{Comparison to Previous Approaches}\label{sec:baselinecomparison}

\begin{figure}[t]
   \centering
   \includegraphics[scale=0.65]{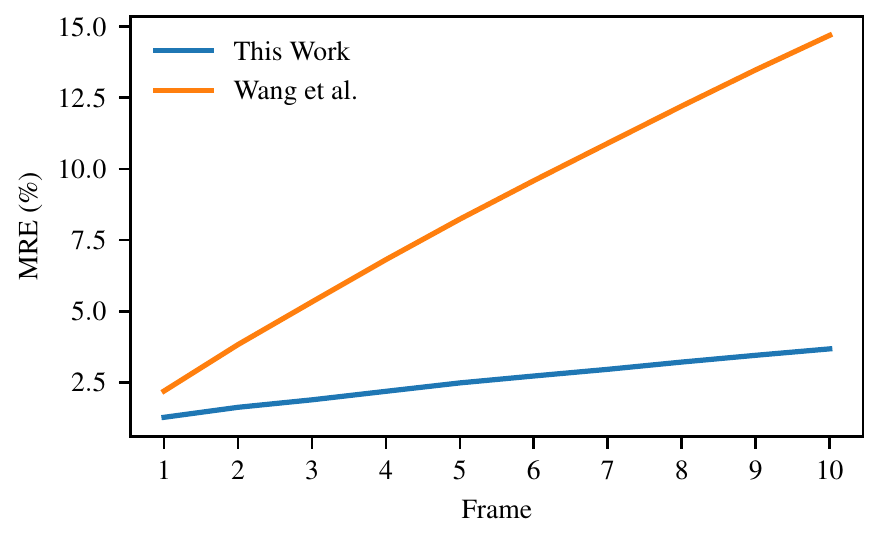}
   \caption{\textbf{Comparison to Depth Transfer Techniques}: Our technique not only remaps the pixels of a previous depth maps but also accounts for the changes in depth.}
   \label{fig:compare_wang}
\end{figure}
\begin{figure*}[t]
   \centering
   \includegraphics{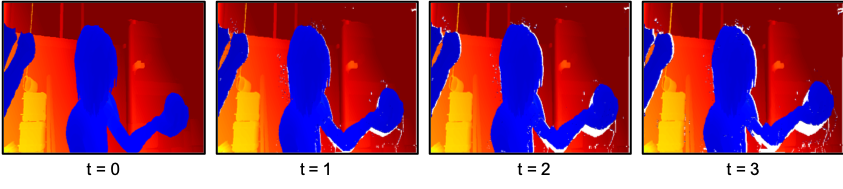}
   \caption{\textbf{Missing Depth}: As the hand moves, part of the background becomes unoccluded, and because its depth was not previously measured, they are missing in the reprojected depth map.}
   \label{fig:unoccluded}
\end{figure*}

\begin{table}[t]
   \centering
   \begin{tabular}{lccccc}\hline
      \toprule
      \tiny{\textbf{Dataset}} & \tiny{\textbf{This Work}} & \tiny{\textbf{Wang \emph{et al.}} \cite{Wang2010}} & \tiny{\textbf{Karsch \emph{et al.}} \cite{Karsch2014}} & \tiny{\textbf{Kumar \emph{et al.}} \cite{Kumar2017}} \\
      \midrule
      DS & 1.8 & 4.6 & 6.9 & 4.8 \\
      MPI & 1.8 & 3.6 & 13.9 & 16.7  \\
      TUM & 3.0 & 6.0 & 18.5 & - \\
      VKITTI & 3.5 & 20.6 & 14.9 & 10.45 \\ 
      \bottomrule
   \end{tabular}
   \caption{\textbf{Comparison to Previous Approaches}: We summarize the MRE of the depth maps estimated by our algorithm and the previous techniques. The code for Kumar \emph{et al.} \cite{Kumar2017} is not publicly available, and we report the results from their paper.}\label{tab:bc}
\end{table}

To better evaluate our approach, we compare its performance to the approaches in Section \ref{sec:relatedworks}. We use the depth transfer methods, namely Wang \emph{et al.} \cite{Wang2010} and Karsch \emph{et al.} \cite{Karsch2014}, because they have the most similar setup. For the non-rigid structure-from-motion techniques, we compare our technique to Kumar \emph{et al.} \cite{Kumar2017}, which estimates dense depth using consecutive frames without segmentation. We do not compare our technique to the neural network-based approaches because these networks are trained on images with different resolutions and characteristics compared to those in the datasets we evaluate on as that would negatively bias its performance.

\subsubsection{Comparison to Depth Transfer Techniques}

We compare our approach to a causal variant of Wang \emph{et al.} \cite{Wang2010}, which only uses the previously measured depth maps to estimate a new one. As stated in Section \ref{sec:relatedworks}, this method would be effective for small changes in depth and in-plane motion. In Figure \ref{fig:compare_wang}, we see that the MRE for Wang \emph{et al.} \cite{Wang2010} increases significantly from frames to frame. This suggests substantial changes in depth in the scenes, which the dense optical flow cannot account for, but is captured by our approach. Consequently, as shown in Table \ref{tab:bc}, our technique outperforms this approach for every dataset we evaluate on.   

Karsch \emph{et al.} \cite{Karsch2014} estimates depth maps by querying from a training set of image and depth map pairs captured from similar scenes. For our experiments, we randomly selected image and depth map pairs from each dataset for this technique to use. We use the default parameters, where each depth map is estimated using 8 examples from the training set. However, even with a training set taken from the same scenes, we see in Table \ref{tab:bc} that our approach outperforms Karsch \emph{et al.} \cite{Karsch2014}. This makes sense because similar images of the same scenes are not guaranteed to have the same depth, and this is reflected by the high MRE.

\subsubsection{Comparison to Non-Rigid Structure-from-Motion}
We also compare our approach to Kumar \emph{et al.} \cite{Kumar2017}, which is similar to our technique in that it estimates depth maps using only two frames and assumes that the scene is locally rigid. This approach is different in that it does not have any previous depth. Therefore in order to estimate depth, it first oversegments the scene into rigid superpixels and then uses the \emph{per-pixel}, dense optical flow to estimate the depth within each superpixel while ensuring scale consistency. This requires restrictive assumptions for each superpixel and is sensitive to the accuracy of the optical flow, which is not always possible to accurately estimate. While this method produces encouraging results (especially since it does not use any previous depth measurements like in our approach and the previous ones), our technique still outperforms it as shown in Table \ref{tab:bc}. This shows that using prior depth allows us to both efficiently and accurately estimate depth in dynamic scenes.

\subsection{Limitations of Our Approach}\label{sec:limitations}
As described in Section \ref{sec:motionestimation}, our algorithm estimates the independent and rigid motions in the scene using the optical flow at a sparse set of key points. Therefore, our approach will naturally fail for scenes with limited texture. Another consequence of this is that our approach also fails to estimate the motion of small rigid segments, where key points are not detected and the optical flow not estimated. One way to detect these scenarios is to examine the photometric error and use it as a measure of confidence for the estimated depth, an area of future exploration.   

Another drawback of our approach stems from the motion assignment and depth map estimation in Section \ref{sec:motionassign}. For regions with limited texture, the pose can be incorrectly assigned and lead to pixels being erroneously reprojected. However, even when the pose is correctly assigned, we still have missing depth estimates in the depth map. Some of these missing pixels arise because reprojecting the neighboring pixels of a rigid segment does not constrain them to be contiguous in the estimated depth map. However, these missing pixels are smaller in area compared to those in regions which were previously occluded, but uncovered due to the motion in the scene. This is shown in Figure \ref{fig:unoccluded}. Furthermore, as the depth maps are sequentially estimated, these holes become more pronounced. While there are many promising infilling approaches \cite{Miao2012,Liu2012,Herrera2013,Qi2013,Buyssens2015}, we avoid them because they often require assumptions that are not geometrically motivated. We can also avoid these infilling techniques because in our problem setup, we can still use the depth sensor. For example, when the depth in a previously occluded region is required by the underlying application, this event can be used to trigger the depth sensor to obtain a new depth map as done in \cite{Noraky2019}.

\section{Conclusion}\label{sec:conclusion}

Depth information is useful for many applications, and active depth sensors are often used to obtain accurate and dense depth measurements. However, for many of these applications, it is undesirable to continuously use the active depth sensor due to issues like its high power consumption while estimating depth using only passively obtained data, like images, have prohibitive computational costs. In this paper, we present a solution that balances the acquisition of depth between the active depth sensor and computation without incurring a large cost. Our algorithm estimates new depth maps using consecutive images and a previously measured depth map for dynamic scenes. Our contribution is to incorporate the previous depth measurements into our formulation, which allows us to efficiently estimate the rigid motions in the scene using sparse optical flow and to accurately assign the estimated rigid motions to the pixels of the previous depth map to obtain a new one. The resulting algorithm is efficient and estimates dense depth maps at up to real-time (30 FPS) on a standard laptop computer. Across different datasets, we show that our technique can reduce the usage of the active depth sensor by over 90\% but still estimate depth maps with an average mean relative error of 2.5\%.

\section*{Acknowledgement}

We thank Analog Devices for funding this research. We also thank the research scientists within the company for helpful discussions and feedback. 

\bibliographystyle{IEEEtran}

\end{document}